\documentclass[twocolumn,english,aps,pra,showpacs]{revtex4}
\usepackage[T1]{fontenc}
\usepackage[latin1]{inputenc}
\usepackage{amsmath}
\usepackage{amssymb}
\usepackage{graphicx}
\usepackage{esint}

\makeatletter
\@ifundefined{textcolor}{}
{%
 \definecolor{BLACK}{gray}{0}
 \definecolor{WHITE}{gray}{1}
 \definecolor{RED}{rgb}{1,0,0}
 \definecolor{GREEN}{rgb}{0,1,0}
 \definecolor{BLUE}{rgb}{0,0,1}
 \definecolor{CYAN}{cmyk}{1,0,0,0}
 \definecolor{MAGENTA}{cmyk}{0,1,0,0}
 \definecolor{YELLOW}{cmyk}{0,0,1,0}
}

\usepackage{babel}

\makeatother

\usepackage{babel}
\begin{document}

\title{Topological superfluid in one-dimensional spin-orbit coupled atomic
Fermi gases}

\author{Xia-Ji Liu$^{1}$ and Hui Hu$^{1}$}

\affiliation{$^{1}$ARC Centre of Excellence for Quantum-Atom Optics,\\
 Centre for Atom Optics and Ultrafast Spectroscopy, \\
 Swinburne University of Technology, Melbourne 3122, Australia}

\date{\today}
\begin{abstract}
We investigate theoretically the prospect of realizing a topological
superfluid in one-dimensional spin-orbit coupled atomic Fermi gases
under Zeeman field in harmonic traps. In the absence of spin-orbit
coupling, it is well-known that the system is either a Bardeen-Cooper-Schrieffer
(BCS) superfluid or an inhomogeneous Fulde-Ferrell-Larkin-Ovchinnikov
(FFLO) superfluid. Here we show that with spin-orbit coupling it could
be driven into a topological superfluid, which supports zero-energy
Majorana modes. However, in the weakly interacting regime the spin-orbit
coupling does not favor the spatially oscillating FFLO order parameter.
As a result, it seems difficult to create an inhomogeneous topological
superfluid in current cold-atom experiments. 
\end{abstract}

\pacs{03.75.Ss, 71.10.Pm, 03.65.Vf, 03.67.Lx}

\maketitle

\section{Introduction}

Topological superfluids are new states of matter that attract intense
attentions in recent years \cite{Qi2010,Hasan2010}. They have a full
pairing gap in the bulk and exotic gapless excitations at the edge
- the so-called Majorana fermions - which obey non-Abelian statistics
\cite{Majorana1937,Wilczek2009}. These excitations are immune to
decoherence caused by local perturbations. By properly braiding excitation
quasiparticles, topological quantum information might be processed.
As a result, topological superfluids could provide an ideal platform
for topological quantum computation \cite{Kitaev2006,Nayak2008}.
Because of this potential application, the realization of topological
superfluids in a well-controlled environment is highly desirable.

Theoretically, there are a number of proposals on realizing a topological
superfluid in two-dimensional (2D) settings, including the use of
2D $p$-wave pairing \cite{Read2000,Mizushima2008}, proximity coupling
to a conventional $s$-wave superconductors for the surface state
of three-dimensional (3D) topological insulators \cite{Fu2008,Sau2010,Alicea2010},
and 2D atomic Fermi gases with strong Rashba spin-orbit coupling \cite{Zhang2008,Liu2012}.
It is also possible to create a topological superfluid in one-dimensional
(1D) solid-state systems by suitably engineering spin-orbit coupling
of electrons, such as InAs wires and banded carbon nanotubes \cite{Lutchyn2010,Oreg2010,Alicea2011,Stoudenmire2011}.
The purpose of this work is to examine the possibility of observing
topological superfluids in 1D ultracold atomic Fermi gases \cite{Jiang2011},
which may be regarded as highly controllable quantum simulators of
the corresponding 1D solid-state systems. We note that 1D atomic Fermi
gases can now be routinely created in cold-atom laboratories \cite{Liao2010}.
The spin-orbit coupling for neutral atoms may also be generated by
using the so-called ``non-Abelian synthetic gauge fields'' technique
\cite{Lin2011,Sau2011}.

Even in the absence of spin-orbit coupling the 1D ultracold atomic
Fermi gas is of great interest. It hosts a Bardeen-Cooper-Schrieffer
(BCS) superfluid and an exotic inhomogeneous Fulde-Ferrell-Larkin-Ovchinnikov
(FFLO) superfluid \cite{Liao2010,Guan2007,Orso2007,Hu2007,Liu2007,Feiguin2007,Tezuka2008,Liu2008},
respectively, in the case of balanced and imbalanced spin-populations.
Here we show that by adding spin-orbit coupling both superfluids can
turn into a topological superfluid. We discuss in detail the resulting
zero-energy Majorana edge modes and their possible experimental signature.
We also explore the possibility of creating an inhomogeneous topological
superfluid with spatially oscillating FFLO order parameter. Unfortunately,
the spin-orbit coupling seems to suppress the FFLO order parameter.
As a result, in the weakly interacting regime we always find the same
topological superfluid with a uniform order parameter, whatever the
initial state is a BCS or FFLO superfluid. Our study is based on the
self-consistent solution of fully microscopic Bogoliubov-de Gennes
(BdG) equations \cite{Liu2007,Liu2008}. It enables ab-initio simulations
under realistic experimental conditions.

The paper is organized as follows. In the next section (Sec. II),
we present the model Hamiltonian and the BdG equations. In Sec. III,
we discuss the phase diagram at a sufficiently large spin-orbit coupling
and the phase transition from BCS superfluid to topological superfluid.
The wave-functions of Majorana edge modes are shown and their possible
experimental detection is considered. In Sec. IV, we present the phase
diagram at a given Zeeman field and show the transition from FFLO
superfluid to topological superfluid. Finally, in Sec. V we provide
conclusions and some final remarks.

\section{Model Hamiltonian and BdG equations}

We consider a trapped two-component 1D atomic Fermi gas under a non-Abelian
gauge field (spin-orbit coupling) and Zeeman field, described by the
model Hamiltonian, 
\begin{eqnarray}
{\cal H} & = & \int dx\psi^{\dagger}\left(x\right)\left[{\cal H}_{0}^{S}\left(x\right)-h\sigma_{z}+\lambda k\sigma_{y}\right]\psi\left(x\right)\nonumber \\
 &  & +g_{1D}\int dx\psi_{\uparrow}^{\dagger}\left(x\right)\psi_{\downarrow}^{\dagger}\left(x\right)\psi_{\downarrow}\left(x\right)\psi_{\uparrow}\left(x\right),\label{Hami}
\end{eqnarray}
 where $\psi^{\dagger}\left(x\right)\equiv[\psi_{\uparrow}^{\dagger}\left(x\right),\psi_{\downarrow}^{\dagger}\left(x\right)]$
denotes collectively the creation field operators for spin-up and
spin-down atoms. In the single-particle Hamiltonian (i.e., the first
line of the above equation), ${\cal H}_{0}^{S}(x)\equiv-(\hbar^{2}/2m)\partial^{2}/\partial x^{2}+m\omega^{2}x^{2}/2-\mu$
describes the single-particle motion in a harmonic trapping potential
$m\omega^{2}x^{2}/2$ and in reference to the chemical potential $\mu$,
the strength of the Zeeman field is denoted by $h$, $\lambda k\sigma_{y}\equiv-i\lambda(\partial/\partial x)\sigma_{y}$
is the spin-orbit coupling term with coupling strength $\lambda$,
$\sigma_{y}$ and $\sigma_{z}$ are the $2\times2$ Pauli matrices.
The second line of the equation is the interaction Hamiltonian, with
the (attractive) interaction strength given by the $s$-wave scattering
length: $g_{1D}=-2\hbar^{2}/(ma_{1D})$.

The model Hamiltonian Eq. (\ref{Hami}) can be realized straightforwardly
with cold fermionic atoms. It is a direct generalization of the standard
model Hamiltonian for a 1D spin-imbalanced Fermi gas, through the
inclusion of a non-Abelian synthetic gauge field $\lambda k\sigma_{y}$.
Experimentally, a bundle of 1D spin-imbalanced atomic Fermi gases
can now be manipulated using 2D optical lattices \cite{Liao2010}.
The generalization of the synthetic gauge field $\lambda k\sigma_{y}$
has already been demonstrated in a 3D Bose gas of $^{87}$Rb atoms
\cite{Lin2011}. In addition, its realization in fermionic atoms has
been proposed \cite{Sau2011}. Therefore, all the techniques required
to simulate Eq. (1) are within current experimental reach.

To understand the 1D superfluidity in the presence of spin-orbit coupling,
we calculate elementary excitations within the mean-field BdG approach
\cite{Liu2007,Liu2008}. The wave-function of low-energy fermionic
quasiparticles $\Psi_{\eta}\left(x\right)$ with energy $E_{\eta}$
is solved by, 
\begin{equation}
{\cal H}_{BdG}\Psi_{\eta}\left(x\right)=E_{\eta}\Psi_{\eta}\left(x\right),\label{BdG}
\end{equation}
 where $\Psi_{\eta}\left(x\right)\equiv[u_{\uparrow\eta}\left(x\right),u_{\downarrow\eta}\left(x\right),v_{\uparrow\eta}\left(x\right),v_{\downarrow\eta}\left(x\right)]^{T}$
in the Nambu spinor representation and the BdG Hamiltonian ${\cal H}_{BdG}$
reads accordingly, \begin{widetext}
\begin{equation}
{\cal H}_{BdG}=\left[\begin{array}{cccc}
{\cal H}_{0}^{S}(x)-h & -\lambda\partial/\partial x & 0 & -\Delta(x)\\
\lambda\partial/\partial x & {\cal H}_{0}^{S}(x)+h & \Delta(x) & 0\\
0 & \Delta^{*}(x) & -{\cal H}_{0}^{S}(x)+h & \lambda\partial/\partial x\\
-\Delta^{*}(x) & 0 & -\lambda\partial/\partial x & -{\cal H}_{0}^{S}(x)-h
\end{array}\right].\label{BdGHami}
\end{equation}
\end{widetext} Here $\Delta(x)=-(g_{1D}/2)\sum_{\eta}[u_{\uparrow\eta}v_{\downarrow\eta}^{*}f(E_{\eta})+u_{\downarrow\eta}v_{\uparrow\eta}^{*}f(-E_{\eta})]$
is the order parameter and $f\left(x\right)\equiv1/[e^{x/(k_{B}T)}+]$
is the Fermi distribution function at temperature $T$. The order
parameter is to be solved self-consistently together with the number
equation for the chemical potential, $\int d{\bf r[}n_{\uparrow}\left({\bf r}\right)+n_{\downarrow}\left({\bf r}\right)]=N$,
where $N$ is the total number of atoms and the density of spin-$\sigma$
atoms is given by, $n_{\sigma}\left(x\right)=(1/2)\sum_{\eta}[\left|u_{\sigma\eta}\right|^{2}f(E_{\eta})+\left|v_{\sigma\eta}\right|^{2}f(-E_{\eta})]$.
We note that the use of Nambu spinor representation leads to an inherent
redundancy built into the BdG Hamiltonian \cite{Hasan2010}. ${\cal H}_{BdG}$
is invariant under the particle-hole transformation: $u_{\sigma}\left(x\right)\rightarrow v_{\sigma}^{*}\left(x\right)$
and $E_{\eta}\rightarrow-E_{\eta}$. Therefore, every eigenstate with
energy $E$ has a partner at $-E$. These two states describe the
same physical degrees of freedom, as the Bogoliubov quasiparticle
operators associated with them satisfy $\Gamma_{E}=\Gamma_{-E}^{\dagger}$.
This redundancy has been removed by multiplying a factor of $1/2$
in the expressions for order parameter and atomic density.

The BdG equation (\ref{BdG}) can be solved by expanding $u_{\sigma\eta}\left(x\right)$
and $v_{\sigma\eta}\left(x\right)$ in the basis of 1D harmonic oscillators.
On such a basis, Eq. (\ref{BdGHami}) is converted to a secular matrix.
A matrix diagonalization then gives the desired quasiparticle energy
spectrum and wave-functions. Numerically, we have to truncate the
summation over the energy levels $\eta$. For this purpose, we adopt
a hybrid strategy developed earlier by us for an imbalanced Fermi
gas without spin-orbit coupling \cite{Liu2007,Liu2008}. We introduce
a high energy cut-off $E_{c}$, above which a local density approximation
(LDA) is used for the high-lying energies and wave-functions. This
leads to an effective coupling constant in the gap equation, $\Delta(x)=-[g_{1D}^{eff}\left(x\right)/2]\sum_{\eta}[u_{\uparrow\eta}v_{\downarrow\eta}^{*}f(E_{\eta})+u_{\downarrow\eta}v_{\uparrow\eta}^{*}f(-E_{\eta})]$,
where $\sum_{\eta}$ is now restricted to $\left|E_{\eta}\right|\leq E_{c}$.
We refer to Ref. \cite{Liu2007} for further details of $g_{1D}^{eff}\left(x\right)$
and the LDA atomic density.

In harmonic traps, it is useful to characterize the interaction strength
by using a dimensionless interaction parameter \cite{Liu2007}, $\gamma\equiv-mg_{1D}/(\hbar^{2}n_{0})=2/(n_{0}a_{1D})$,
where $n_{0}$ is the zero-temperature center density of an ideal
two-component Fermi gas with equal spin populations $N/2$. In the
Thomas-Fermi approximation (or LDA), $n_{0}=2N^{1/2}/(\pi a_{ho})$
and $a_{ho}=\sqrt{\hbar/(m\omega)}$ is the characteristic oscillator
length of the trap. Therefore, the dimensionless interaction parameter
is given by, 
\begin{equation}
\gamma=\frac{1}{\pi N^{1/2}}\left(\frac{a_{ho}}{a_{1D}}\right).
\end{equation}
 We note that, for a 1D atomic Fermi gas created using 2D optical
lattices, the typical dimensionless interaction strength is about
$\gamma=3\sim5$ \cite{Liao2010,Liu2007}. Throughout the paper, we
shall take a slightly smaller value of $\gamma=\pi/2\simeq1.6$, in
order to validate the mean-field treatment. It is also convenient
to use the Thomas-Fermi energy $E_{F}=(N/2)\hbar\omega$ and Thomas-Fermi
radius $x_{F}=N^{1/2}a_{ho}$ as the units for energy and length,
respectively. For the spin-orbit coupling, we use a dimensionless
parameter $\lambda k_{F}/E_{F}$, where $k_{F}=\sqrt{2mE_{F}}$ is
the Thomas-Fermi wavevector. We have performed numerical calculations
for a Fermi gas of $N=100$ fermions in traps at both zero temperature
and finite temperature. In the following, we present only the zero-temperature
results, as the inclusion of a finite but small temperature (i.e.,
$T=0.1T_{F}$) essentially does not affect the results. The Fermi
energy is $E_{F}=(N/2)\hbar\omega=50\hbar\omega$. We have taken a
cut-off energy $E_{c}=4E_{F}=200\hbar\omega$ and have used $3N=300$
1D harmonic oscillators as the expansion functions. These parameters
are already sufficiently large to ensure the accuracy of calculations.

\section{Phase diagram at a given spin-orbit coupling}

The most salient feature of a spin-orbit coupled Fermi gas is the
appearance of topological superfluidity and zero-energy Majorana fermion
mode, under an appropriate Zeeman field. The quasiparticle operators
of Majorana fermions are real and satisfy $\gamma=\gamma^{\dagger}$,
which means that a quasiparticle is its own antiparticle \cite{Majorana1937,Wilczek2009}.
Mathematically, we can always write a complex ordinary fermion operator
$c$ in terms of two real Majorana fermions $\gamma_{1}$ and $\gamma_{2}$,
such as $c=\gamma_{1}-i\gamma_{2}$. An ordinary fermion may therefore
be viewed as a bound state of two Majorana fermions, which in general
can not be deconfined. However, the deconfinement does happen in a
topological superfluid, leading to two Majorana fermions localized
respectively at the two edges of topological superfluid. This can
be clearly seen with the help of the particle-hole redundancy of the
BdG equation \cite{Hasan2010,Liu2012}. Let us image that we have
a zero-energy solution $E=0$. Because of the particle-hole redundancy
$\Gamma_{E}=\Gamma_{-E}^{\dagger}$, we will immediately have $\Gamma_{0}=\Gamma_{0}^{\dagger}$
- exactly the defining feature of a Majorana fermion. We note that,
zero-energy Majorana fermions should always come in pairs, since the
original model Hamiltonian describes ordinary fermions only and each
Majorana fermion is just a half of ordinary fermion. It is straightforward
to check from the BdG Hamiltonian that the wave functions of two paired
Majorana fermions should satisfy $u_{\sigma}\left(x\right)=v_{\sigma}^{*}\left(x\right)$
and $u_{\sigma}\left(x\right)=-v_{\sigma}^{*}\left(x\right)$, respectively.
The former follows the particle-hole symmetry, while the later is
required to express an ordinary fermion by two Majorana fermions %
\footnote{The ordinary fermion operator at $E=0$ is given by $c=\Gamma_{0}+\tilde{\Gamma}_{0}$.
By defining Majorana operators $\gamma_{1}=\Gamma_{0}$ and $\gamma_{2}=i\tilde{\Gamma}_{0}$,
we express $c=\gamma_{1}-i\gamma_{2}$, as anticipated. For $\tilde{\Gamma}_{0}$,
we must have $\tilde{\Gamma}_{0}=-\tilde{\Gamma}_{0}^{\dagger}$.
The associated wave-functions satisfy $u_{\sigma}\left(x\right)=-v_{\sigma}^{*}\left(x\right)$.%
}.

To satisfy the prescription of a zero-energy solution for a topological
superfluid, the quasiparticle energy spectrum must become gapless
at a certain point. In the case of a {\em homogeneous }spin-orbit
coupled Fermi gas under a Zeeman field, this happens at a critical
Zeeman field \cite{Lutchyn2010,Oreg2010}, 
\begin{equation}
h_{c}=\sqrt{\mu^{2}+\Delta^{2}}.
\end{equation}
 The system will be in a conventional superfluid at $h<h_{c}$ and
in a topological superfluid at $h>h_{c}$. For a {\em trapped}
system, however, the critical Zeeman field may become position dependent.
As a result, in harmonic traps we would have a mixed phase with both
conventional and topological superfluid components, which separate
spatially in real space. Without confusion, we shall still refer to
such a mixed phase as a topological superfluid.

\subsection{Phase diagram at $\lambda k_{F}/E_{F}=1$}

\begin{figure}[htp]
\begin{centering}
\includegraphics[clip,width=0.45\textwidth]{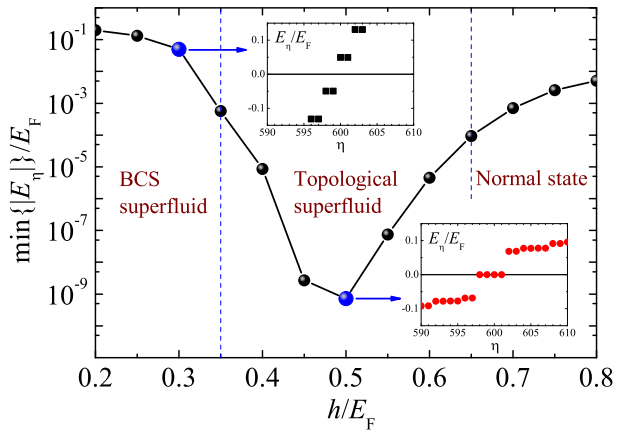} 
\par\end{centering}

\caption{(color online) Phase diagram at a given spin-orbit coupling $\lambda k_{F}/E_{F}=1$,
determined from the behavior of the lowest eigenenergy of Bogoliubov
quasiparticle spectrum, $\min\{\left|E_{\eta}\right|\}$. As the Zeeman
field increases, the system evolves from a conventional BCS superfluid
to a topological superfluid, and finally to a normal state. The two
insets in the middle and right show the quasiparticle spectrum at
$h/E_{F}=0.3$ and $0.5$, respectively.}

\label{fig1} 
\end{figure}

In Fig. 1 we report the phase diagram at a fixed spin-orbit coupling
strength $\lambda k_{F}/E_{F}=1$. The emergence of a topological
superfluid can be clearly revealed by the behavior of the lowest eigenenergy
of the quasiparticle energy spectrum. As shown in the middle inset,
at a small Zeeman field the energy spectrum is gapped. However, by
increasing the Zeeman field above a critical value of $h\sim0.35E_{F}$,
the lowest eigenenergy becomes exponentially small. Four quasiparticle
modes with nearly zero energy appear, as seen clearly from the right
inset. By further increasing the Zeeman field ($h>0.65E_{F}$), the
system will be driven into a normal state with negligible superfluid
order parameter.

\begin{figure}[htp]
\begin{centering}
\includegraphics[clip,width=0.45\textwidth]{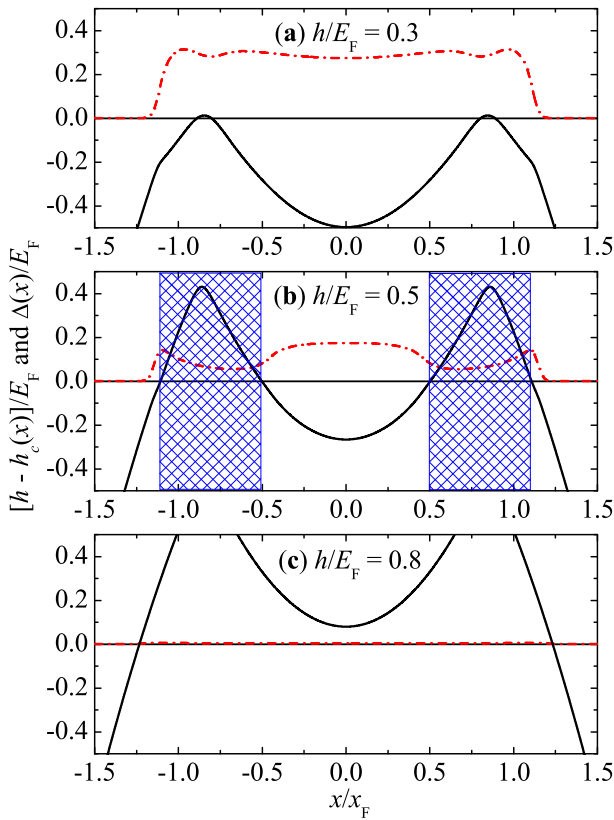} 
\par\end{centering}

\caption{(color online) Spatial dependence of the critical Zeeman field $h-h_{c}(x)$
(solid lines) and the superfluid order parameter $\Delta(x)$ (dot-dashed
lines), at $\lambda k_{F}/E_{F}=1$ and at three Zeeman fields $h/E_{F}=0.3$,
$0.5$, and $0.8$. The cross-patterns highlight the area in which
the atoms are in the topological superfluid state.}

\label{fig2} 
\end{figure}

The appearance of the topological superfluid can also be monitored
by the calculation of $h-h_{c}(x)$, where $h_{c}(x)=\sqrt{\mu^{2}(x)+\Delta^{2}(x)}$
is the local critical Zeeman field for a local uniform cell at position
$x$ with the local chemical potential $\mu(x)\equiv\mu-m\omega^{2}x^{2}/2$
and order parameter $\Delta(x)$. The local uniform cell would be
in the topological superfluid state if $h>h_{c}(x)$. In Fig. 2, we
present $h-h_{c}(x)$ and $\Delta(x)$ at different phases. In accord
with Fig. 1, at a small field $h=0.3E_{F}$ (Fig. 2(a)), $h<h_{c}(x)$
for any position $x$ and the whole Fermi cloud is in the conventional
superfluid. At the field $h=0.5E_{F}$ (Fig. 2(b)), we find $h>h_{c}(x)$
at the two wings of the harmonic trap and therefore there are two
blocks of topological superfluid, as highlighted by the cross-pattern.
At an even large Zeeman-field (Fig. 2(c)), the area of $h>h_{c}(x)$
extends over the whole system. However, the superfluid order parameter
becomes so small, the system can no longer be viewed a superfluid.
We note that, at large attractive interactions where the order parameter
is not destroyed by large Zeeman field, it is possible to have a single
topological superfluid throughout the whole Fermi cloud.

\subsection{Majorana fermions}

\begin{figure}[htp]
\begin{centering}
\includegraphics[clip,width=0.45\textwidth]{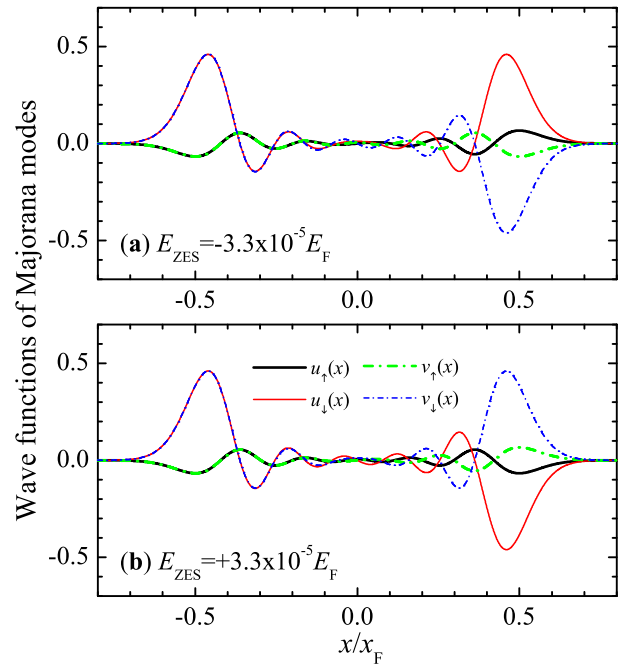} 
\par\end{centering}

\caption{(color online) Wave functions of the paired Majorana modes at the
inner wing of the trap, $x\simeq\pm0.5x_{F}$: one has the energy
$E_{ZES}\simeq-3.3\times10^{-5}E_{F}$ (a) and the other $E_{ZES}\simeq+3.3\times10^{-5}E_{F}$
(b). Both modes satisfy the symmetry requirement for Majorana wave-functions.
The wave functions are in units of $a_{ho}^{-1/2}$. Here $h=0.5E_{F}$
and $\lambda k_{F}/E_{F}=1$. The system is in the topological superfluid
phase.}

\label{fig3} 
\end{figure}

\begin{figure}[htp]
\begin{centering}
\includegraphics[clip,width=0.45\textwidth]{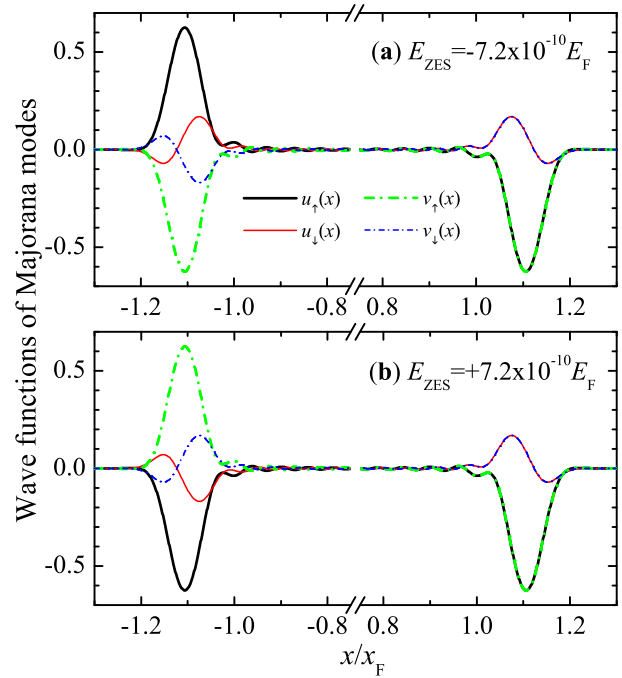} 
\par\end{centering}

\caption{(color online) Wave functions of the paired Majorana modes at the
outer wing of the trap, $x\simeq\pm1.1x_{F}$: one has the energy
$E_{ZES}\simeq-7.2\times10^{-10}E_{F}$ (a) and the other $E_{ZES}\simeq+7.2\times10^{-10}E_{F}$
(b). Other parameters are the same as in Fig. 3.}

\label{fig4} 
\end{figure}

In {\em each} of the topological superfluid phases, we should find
two Majorana fermion modes, well-localized at the two edges respectively.
At the Zeeman field $h=0.5E_{F}$, we therefore could have four Majorana
fermions, as indicated by the energy spectrum in the right inset of
Fig. 1. The wave functions of these Majorana fermions are shown in
Figs. 3 and 4 for states localized at $x\simeq\pm0.5x_{F}$ and $\pm1.1x_{F}$,
respectively. It is interesting that the wave functions of two {\em
paired} Majorana fermions, for example, these located at $x\simeq-0.5x_{F}$
and $x\simeq+0.5x_{F}$ (Fig. 3), tend to interfere with each other
\cite{Liu2012,Mizushima2010}. This quasiparticle interference or
tunneling leads to the splitting of degenerate zero energy Majorana
modes to a finite but exponentially small energy: $E_{ZES}\simeq\pm3.3\times10^{-5}E_{F}$.
The tunneling between the paired Majorana fermions at the outer wing
of the trap, $x\simeq\pm1.1x_{F}$, is more difficult (see Fig. 4),
so the energy splitting is much smaller, i.e., $E_{ZES}\simeq\pm7.2\times10^{-10}E_{F}$.
It is readily seen that the paired wave functions satisfy either $u_{\sigma}\left(x\right)=v_{\sigma}^{*}\left(x\right)$
or $u_{\sigma}\left(x\right)=-v_{\sigma}^{*}\left(x\right)$, as anticipated
by the required symmetry of Majorana wave functions.

\subsection{Density distribution and local density of states}

We now consider the possible experimental signature for observing
topological superfluid and the associated Majorana fermions. The useful
experimental tools include in-situ absorption imaging and spatially
resolved radio-frequency (rf) spectroscopy \cite{Shin2007}, which
give respectively the density distribution and the local density of
states of the Fermi cloud \cite{coldatomSTM}.

\begin{figure}[htp]
\begin{centering}
\includegraphics[clip,width=0.45\textwidth]{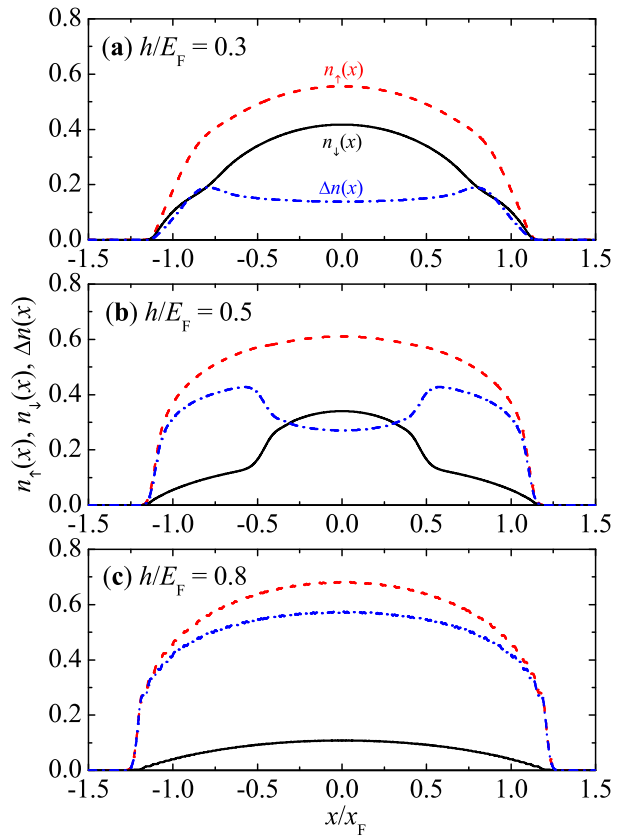} 
\par\end{centering}

\caption{(color online) The spin-up and spin-down density distribution, $n_{\uparrow}(x)$
(dashed lines) and $n_{\uparrow}(x)$ (solid lines), and their difference
$\Delta n(x)=n_{\uparrow}(x)-n_{\downarrow}(x)$ (dot-dashed lines),
are shown at the conventional superfluid phase (a), topological superfluid
state (b), and normal state (c). The density distributions are in
units of the Thomas-Fermi density $n_{0}=2N^{1/2}/(\pi a_{ho})$.
The spin-orbit coupling is $\lambda k_{F}/E_{F}=1$.}

\label{fig5} 
\end{figure}

In Fig. 5, we plot the spin-up $n_{\uparrow}(x)$ and spin-down $n_{\downarrow}(x)$
density distribution and their difference $\Delta n(x)=n_{\uparrow}(x)-n_{\downarrow}(x)$
at different phases. While the shape of the spin-up density distribution
$n_{\uparrow}(x)$ is nearly unchanged across different phases, in
the topological superfluid phase (see Fig. 5(b) at $h=0.5E_{F}$)
the spin-down density distribution $n_{\downarrow}(x)$ shows an interesting
bi-modal structure. It decreases rapidly when the atoms enter the
topological area from the center. Accordingly, a broad dip appear
in the density difference around the trap center. The bi-modal distribution
in $n_{\downarrow}(x)$ may be regarded as a useful and convenient
feature to identify the topological superfluid. However, it is not
a characteristic feature for identifying the Majorana modes, as the
contribution of the Majorana modes to the density distribution is
negligibly small, i.e., relatively at the order of $N^{-1/2}$.

\begin{figure}[htp]
\begin{centering}
\includegraphics[clip,width=0.45\textwidth]{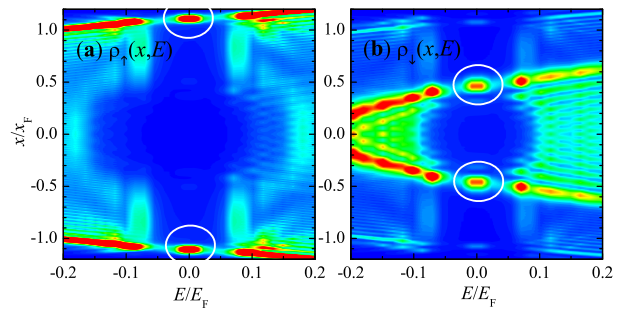} 
\par\end{centering}

\caption{(color online) Linear contour plot (in arbitrary units) for the local
density of states of spin-up atoms $\rho_{\uparrow}(x,E)$ (a) and
of spin-down atoms $\rho_{\downarrow}(x,E)$ (b). The contributions
from Majorana fermions are highlighted by circles. Here the Fermi
cloud is in the topological superfluid state with parameters $h=0.5E_{F}$
and $\lambda k_{F}/E_{F}=1$. In the calculation, the $\delta$-function
in $\rho_{\sigma}(x,E)$ has been simulated by a Lorentz distribution
with a small energy broadening $\Gamma=0.01E_{F}$.}

\label{fig6} 
\end{figure}

A practical way to probe the Majorana fermions is to measure the local
density of states using the spatially resolved rf spectroscopy \cite{Shin2007,coldatomSTM},
with which we anticipate that the contributions of Majorana fermions
will be well-isolated in both energy domain and real space. The local
density of states for spin-up and spin-down atoms is defined by, 
\begin{equation}
\rho_{\sigma}(x,E)=\frac{1}{2}\sum_{\eta}\left[\left|u_{\sigma\eta}\right|^{2}\delta\left(E-E_{\eta}\right)+\left|v_{\sigma\eta}\right|^{2}\delta\left(E+E_{\eta}\right)\right].
\end{equation}
 In Fig. 6, we report the local density of states in the topological
superfluid state. Near the zero energy, the contributions from Majorana
fermions are clearly visible and are well-separated from other quasiparticle
contributions by an energy gap $\Delta\sim0.1E_{F}$. It is interesting
to note that the Majorana modes at $x\simeq\pm1.1x_{F}$ and $\pm0.5x_{F}$
contribute to $\rho_{\uparrow}(x,E)$ and $\rho_{\downarrow}(x,E)$,
respectively. This can be understood from the wave function of Majorana
modes, as shown in Figs. 3 and 4. The wave-functions at $\pm0.5x_{F}$
are dominated by the spin-down component, while the wave-functions
at $\pm1.1x_{F}$ have mainly the spin-up component.

\section{Phase diagram at a given Zeeman field}

We now turn to consider the possibility of observing a topological
superfluid with spatially oscillating order parameter \cite{Hu2007,Liu2007}.
In the absence of spin-orbit coupling, it is known that the ground
state of an imbalance 1D Fermi gas under Zeeman field can be an inhomogeneous
FFLO superfluid with oscillating order parameter. It is therefore
natural to ask: what is the fate of such a FFLO superfluid when we
switch on the spin-orbit coupling?

\begin{figure}[htp]
\begin{centering}
\includegraphics[clip,width=0.45\textwidth]{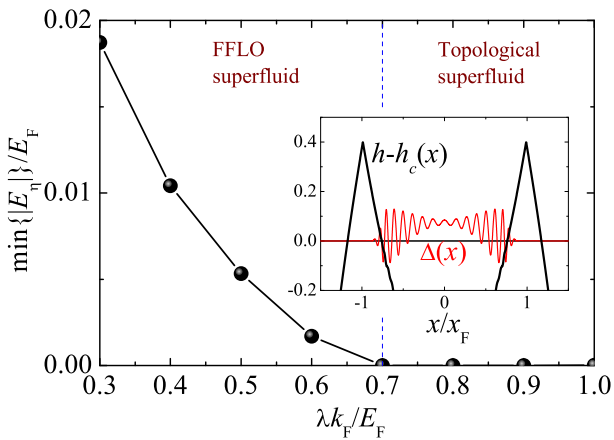} 
\par\end{centering}

\caption{(color online) Phase diagram at a given Zeeman field $h=0.4E_{F}$,
determined from the behavior of the lowest eigenenergy of Bogoliubov
quasiparticle spectrum, $\min\{\left|E_{\eta}\right|\}$. As the spin-orbit
coupling increases, the system evolves from a FFLO superfluid to a
topological superfluid. The inset shows the critical Zeeman field
$h-h_{c}(x)$ and the order parameter $\Delta(x)$ at $\lambda k_{F}/E_{F}=0.3$,
where the Fermi gas is in the FFLO superfluid state.}

\label{fig7} 
\end{figure}

\begin{figure}[htp]
\begin{centering}
\includegraphics[clip,width=0.45\textwidth]{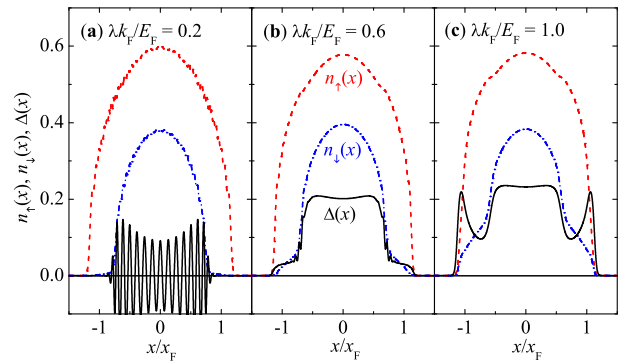} 
\par\end{centering}

\caption{(color online) Density distributions and order parameter at $h=0.4E_{F}$
and at three different spin-orbit couplings: $\lambda k_{F}/E_{F}=0.2$
(a), $0.6$ (b), and $1.0$ (c). The density distribution $n_{\sigma}(x)$
is in units of the Thomas-Fermi density $n_{0}=2N^{1/2}/(\pi a_{ho})$.
The order parameter $\Delta(x)$ is in units of the Fermi energy $E_{F}$.}

\label{fig8} 
\end{figure}

In Fig. 7, we present the phase diagram at a given Zeeman field $h=0.4E_{F}$,
determined again by tracing the behavior of the lowest eigenenergy
of the quasiparticle spectrum as a function of the spin-orbit coupling.
The density distributions and order parameter are reported in Fig.
8 for three values of spin-orbit coupling. At small spin-orbit coupling,
we find a stable FFLO order parameter which is modified slightly by
the spin-orbit coupling. However, in the area where $\Delta_{FFLO}\left(x\right)$
is nonzero, the criterion for a topological superfluid $h>h_{c}(x)$
is always not satisfied, as seen from the inset for the case of $\lambda k_{F}/E_{F}=0.3$.
This excludes the coexistence of FFLO superfluid and topological order.
As a result, the energy spectrum is gapped and $\min\{\left|E_{\eta}\right|\}>0$.
With increasing the spin-orbit coupling above $\lambda k_{F}/E_{F}\simeq0.6$,
we observe that the lowest eigenenergy becomes exponentially small,
suggesting a topological superfluid. However, in this case, the order
parameter no longer oscillates in real space, as shown in Figs. 8(b)
and 8(c). Therefore, we conclude that it seems impossible to create
an inhomogeneous topological superfluid with spatially oscillating
order parameter in 1D spin-orbit coupled Fermi gas, if we do not tailor
specifically the geometry or other parameters of the Fermi cloud.

\section{Conclusions}

In conclusions, we have investigated theoretically the properties
of a 1D imbalanced Fermi gas under non-Abelian synthetic gauge field.
We have predicted that by suitably tuning the strength of spin-orbit
coupling and Zeeman field, it is possible to create a topological
superfluid, which hosts Majorana zero-energy fermions at its edge.
The order parameter in the topological superfluid is always of the
conventional Bardeen-Cooper-Schrieffer type, as the spin-orbit coupling
tends to destroy inhomogeneous Fulde-Ferrell-Larkin-Ovchinnikov pairing.
Experimentally, the topological superfluid may be identified from
the bimodal distribution of the spin-down atomic density by using
in-situ absorption imaging. The associated Majorana fermions may be
detected by applying the spatially resolved radio-frequency spectroscopy,
which would show a well-isolated signal at zero energy.

At the end of this paper, we would like to emphasize that the ultracold
atomic Fermi gas with non-Abelian synthetic gauge field is an ideal
platform for creating topological superfluid and manipulating Majorana
fermions, because of its unprecedented controllability and flexibility.
This system can now be readily realized in ultracold atom laboratories.

\section*{Acknowledgments}

This work was supported by the ARC Discovery Project (Grant No. DP0984637
and DP0984522) and NFRP-China (Grant No. 2011CB921502).


\begin{thebibliography}{10}
\bibitem{Qi2010} X.-L. Qi and S.-C. Zhang, Physics Today \textbf{63},
33 (2010).

\bibitem{Hasan2010} M. Z. Hasan and C. L. Kane, Rev. Mod. Phys. \textbf{82},
3045 (2010).

\bibitem{Majorana1937} E. Majorana, Nuovo Cimennto \textbf{14}, 171
(1937).

\bibitem{Wilczek2009} F. Wilczek, Nature Phys. \textbf{5}, 614 (2009).

\bibitem{Kitaev2006} A. Kitaev, Ann. Phys. (N.Y.) \textbf{321}, 2
(2006).

\bibitem{Nayak2008} C. Nayak, S. Simon, A. Stern, M. Freedman, and
S. Das Sarma, Rev. Mod. Phys. \textbf{80}, 1083 (2008).

\bibitem{Read2000} N. Read and D. Green, Phys. Rev. B \textbf{61},
10267 (2000).

\bibitem{Mizushima2008} T. Mizushima, M. Ichioka, and K. Machida,
Phys. Rev. Lett. \textbf{101}, 150409 (2008).

\bibitem{Fu2008} L. Fu and C. L. Kane, Phys. Rev. Lett. \textbf{100},
096407 (2008).

\bibitem{Sau2010} J. D. Sau, R. M. Lutchyn, S. Tewari, and S. Das
Sarma, Phys. Rev. Lett. \textbf{104}, 040502 (2010).

\bibitem{Alicea2010} J. Alicea, Phys. Rev. B \textbf{81}, 125318
(2010).

\bibitem{Zhang2008} C. Zhang, S. Tewari, R. Lutchyn, and S. Das Sarma,
Phys. Rev. Lett. \textbf{101}, 160401 (2008).

\bibitem{Liu2012} X.-J. Liu, L. Jiang, H. Pu, and H. Hu, eprint arXiv:1111.1798.

\bibitem{Lutchyn2010} R. M. Lutchyn, J. D. Sau, and S. D. Sarma,
Phys. Rev. Lett. \textbf{105}, 077001 (2010).

\bibitem{Oreg2010} Y. Oreg. G. Refael, and F. von Oppen, Phys. Rev.
Lett. \textbf{105}, 177002 (2010).

\bibitem{Alicea2011} J. Alicea, Y. Oreg, G. Refael, F. von Oppen,
and M. P. A. Fisher, Nature Phys. \textbf{7}, 412 (2011).

\bibitem{Stoudenmire2011} E. M. Stoudenmire, J. Alicea, O. A. Starykh,
and M. P.A. Fisher, Phys. Rev. B \textbf{84}, 014503 (2011).

\bibitem{Jiang2011} L. Jiang, T. Kitagawa, J. Alicea, A. R. Akhmerov,
D. Pekker, G. Refael, J. I. Cirac, E. Demler3, M. D. Lukin, and P.
Zoller, Phys. Rev. Lett. \textbf{106}, 220402 (2011).

\bibitem{Liao2010} Y. A. Liao, A. S. C. Rittner, T. Paprotta, W.
Li, G. B. Partridge, R. G. Hulet, S. K. Baur, and E. J. Mueller, Nature
(London) \textbf{467}, 567 (2010).

\bibitem{Lin2011} Y.-J. Lin, K. Jiménez-Garc\'{i}a, and I. B. Spielman,
Nature (London) \textbf{471}, 83 (2011).

\bibitem{Sau2011} J. D. Sau, R. Sensarma, S. Powell, I. B. Spielman,
and S. Das Sarma, Phys. Rev. B \textbf{83}, 140510(R) (2011).

\bibitem{Guan2007} X.-W. Guan, M. T. Batchelor, C. Lee, and M. Bortz,
Phys. Rev. B \textbf{76}, 085120 (2007).

\bibitem{Orso2007} G. Orso, Phys. Rev. Lett. \textbf{98}, 070402
(2007).

\bibitem{Hu2007} H. Hu, X.-J. Liu, and P. D. Drummond, Phys. Rev.
Lett. \textbf{98}, 070403 (2007).

\bibitem{Liu2007} X.-J. Liu, H. Hu, and P. D. Drummond, Phys. Rev.
A \textbf{76}, 043605 (2007).

\bibitem{Feiguin2007} A. E. Feiguin and F. Heidrich-Meisner, Phys.
Rev. B \textbf{76}, 220508(R) (2007).

\bibitem{Tezuka2008} M. Tezuka and M. Ueda, Phys. Rev. Lett. \textbf{100},
110403 (2008).

\bibitem{Liu2008} X.-J. Liu, H. Hu, and P. D. Drummond, Phys. Rev.
A \textbf{78}, 023601 (2008).

\bibitem{Mizushima2010} T. Mizushima, K. Machida, Phys. Rev. A \textbf{81},
053605 (2010).

\bibitem{Shin2007} Y. Shin, C. H. Schunck, A. Schirotzek, and W.
Ketterle, Phys. Rev. Lett. \textbf{99}, 090403 (2007).

\bibitem{coldatomSTM} L. Jiang, L. O. Baksmaty, H. Hu, Y. Chen, and
H. Pu, Phys. Rev. A \textbf{83}, 061604(R) (2011).\end{thebibliography}
\end{document}